\documentclass[twocolumn,showpacs,aps,floatfix,prd,nofootinbib,superscriptaddress]{revtex4}

\usepackage{graphicx}
\usepackage{dcolumn}
\usepackage{amsmath}
\usepackage{epsfig}
\RequirePackage{xspace} 

\newcommand{\BABARPubYear}    {08}
\newcommand{\BABARPubNumber}  {024}
\newcommand{\SLACPubNumber}   {13338}

\usepackage{relsize}
\def\babar{\mbox{\slshape B\kern-0.1em{\smaller A}\kern-0.1em
    B\kern-0.1em{\smaller A\kern-0.2em R}}}
\def\pep {PEP-II}
\def\BF{$B$ Factory}

\def\mes {\ensuremath{m_{\mbox{\scriptsize ES}} }\xspace}
\def\de   {\ensuremath{\Delta E}\xspace}
\def\GeV{\;\mbox{GeV}}
\def\MeV{\;\mbox{MeV}}
\def\MeVc{\;\mbox{MeV}/c}
\def\GeVc{\;\mbox{GeV}/c}
\def\GeVcc{\;\mbox{GeV}/c^2}
\def\MeVcc{\;\mbox{MeV}/c^2}
\mathchardef\Upsilon="7107
\def\Y#1S{\ensuremath{\Upsilon{(#1S)}}\xspace}
\def\FourS {\Y4S}
\newcommand{\dedx}{\ensuremath{\mathrm{d}\hspace{-0.1em}E/\mathrm{d}x}\xspace}

\def\B       {\ensuremath{B}\xspace}
\def\Bbar    {\kern 0.18em\overline{\kern -0.18em B}{}\xspace}
\def\BB      {\ensuremath{B\Bbar}\xspace} 
\def\Bz      {\ensuremath{B^0}\xspace}
\def\Bzb     {\ensuremath{\Bbar^0}\xspace}
\def\BzBzb   {\ensuremath{\Bz {\kern -0.16em \Bzb}}\xspace}
\def\Bu      {\ensuremath{B^+}\xspace}
\def\Bub     {\ensuremath{B^-}\xspace}
\def\Bp      {\ensuremath{\Bu}\xspace}

\def\BpBm    {\ensuremath{\Bp {\kern -0.16em \Bub}}\xspace}
\def\Bs      {\ensuremath{B_s}\xspace}
\def\BR      {{\ensuremath{\cal B}\xspace}}
\def\L       {{\ensuremath{\cal LR}\xspace}}
\def\epem    {\ensuremath{e^+e^-}\xspace}
\def\pip     {\ensuremath{\pi^+}\xspace}
\def\pim     {\ensuremath{\pi^-}\xspace}
\def\pipm    {\ensuremath{\pi^\pm}\xspace}

\def\piz     {\ensuremath{\pi^0}\xspace}
\def\etaz    {\ensuremath{\eta}\xspace}
\def\rhoz    {\ensuremath{\rho^0}\xspace}
\def\rhop    {\ensuremath{\rho^+}\xspace}
\def\en         {\ensuremath{e^-}\xspace}   
\def\ep         {\ensuremath{e^+}\xspace}
\def\q      {\ensuremath{q}\xspace}

\def\qqbar  {\ensuremath{q\overline q}\xspace}

\def\Kstar   {\ensuremath{K^*}\xspace}
\def\u     {\ensuremath{u}\xspace}
\def\d     {\ensuremath{d}\xspace}
\def\s     {\ensuremath{s}\xspace}
\def\c     {\ensuremath{c}\xspace}
\def\g     {\ensuremath{\gamma}\xspace}
\def\degrees{\ensuremath{^{\circ}}\xspace}

\newcommand{\epjBase}        {Eur.\ Phys.\ Jour.\xspace}
\newcommand{\jprlBase}       {Phys.\ Rev.\ Lett.\xspace}
\newcommand{\jprBase}        {Phys.\ Rev.\xspace}
\newcommand{\jplBase}        {Phys.\ Lett.\xspace}
\newcommand{\npBase}         {Nucl.\ Phys.\xspace}
\newcommand{\zpBase}         {Z.\ Phys.\xspace}
\newcommand{\nimBaseA}       {Nucl.\ Instrum.\ Methods Phys.\ Res., Sect.\ A\xspace}
\newcommand{\jprl}      [1]  {\jprlBase\ {\bf #1}}
\newcommand{\jprd}      [1]  {\jprBase\ D~{\bf #1}}
\newcommand{\plb}       [1]  {\jplBase\ B~{\bf #1}}
\newcommand{\epjc}      [1]  {\epjBase\ C~{\bf #1}}

\newcommand{\npb}       [1]  {\npBase\ B~{\bf #1}}

\newcommand{\jhep}      [1]  {{J.\ High.\ Energy\ Phys.\ {\bf #1}}}
\newcommand{\jpg}       [1]  {{J.\ Phys.\ {\bf G{\bf #1}}}}

\newcommand{\ml}        [1]  {{Mach.\ Learn.\ {\bf #1}}} 
\newcommand{\zpc}       [1]  {\zpBase\ C~{\bf #1}}
\newcommand{\nima}      [1]  {\nimBaseA~{\bf #1}}
\newcommand{\gev}            {\ensuremath{\mathrm{\,Ge\kern -0.1em V}}\xspace}
\newcommand{\mev}            {\ensuremath{\mathrm{\,Me\kern -0.1em V}}\xspace}

\def\figurebox#1#2#3{%
    \def\arg{#3}%
    \ifx\arg\empty
    {\hfill\vbox{\hsize#2\hrule\hbox to #2{\vrule\hfill\vbox to #1{\hsize#2\vfill}\vrule}\hrule}\hfill}%
    \else
    {\hfill\epsfbox{#3}\hfill}%
    \fi}

\def\brpg      {\ensuremath{\Bp \to \rho^+\gamma}\xspace}
\def\brzg      {\ensuremath{\Bz \to \rho^0\gamma}\xspace}
\def\bomg      {\ensuremath{\Bz \to \omega\gamma}\xspace}
\def\avbr      {\ensuremath{{\BR}[B \to (\rho/\omega)\gamma]}\xspace}
\def\VtdVts    {\ensuremath{|V_{td}/V_{ts}|}\xspace}
\def\bdg       {\ensuremath{b \to d \gamma}\xspace}

\def\bkg       {\ensuremath{B \to \Kstar\gamma}\xspace}
\def\brg       {\ensuremath{B \to \rho\gamma}\xspace}
\def\brog       {\ensuremath{B \to (\rho/\omega)\gamma}\xspace}
\def\Bd        {\ensuremath{B_d}\xspace}
\def\rppipi    {\ensuremath{\rho^0\to\pip\pim}\xspace}
\def\rzpipi    {\ensuremath{\rho^+\to\pip\piz}\xspace}
\def\ompipi    {\ensuremath{\omega\to\pip\pim\piz}\xspace}

\def\bkpg      {\ensuremath{B^+ \to K^{*+}\gamma}\xspace}
\def\bkzg      {\ensuremath{B^0 \to K^{*0}\gamma}\xspace}
\def\brpiz     {\ensuremath{B^+ \to \rho^+\pi^0}\xspace}
\def\rppiz       {\ensuremath{\rho^+\pi^0}\xspace}
\def\bromg     {\ensuremath{B \to (\rho/\omega)\gamma}\xspace}
\def\broomg     {\ensuremath{B \to \rho(\omega)\gamma}\xspace}
\def\brppiz    {\ensuremath{B^+ \to \rho^+\piz}\xspace}

\def\lumBB     {\ensuremath{423}\xspace}
\def\numBB     {\ensuremath{465}\xspace}
\def\valBFbase {\ensuremath{\times{10^{-6}}}\xspace}
\def\valBFrp   {\ensuremath{1.20^{+0.42}_{-0.37}\pm 0.20}\xspace} 
\def\valBFrz   {\ensuremath{0.97^{+0.24}_{-0.22}\pm 0.06}\xspace} 
\def\valBFom   {\ensuremath{0.50^{+0.27}_{-0.23}\pm 0.09}\xspace} 
\def\valBFomUL {\ensuremath{0.9}\xspace} 
\def\valBFav   {\ensuremath{1.63^{+0.30}_{-0.28}\pm 0.16}\xspace} 
\def\valBFavrho   {\ensuremath{1.73^{+0.34}_{-0.32}\pm0.17}\xspace} 
\def\effrp     {\ensuremath{4.2}}
\def\effrz     {\ensuremath{7.7}}
\def\effom     {\ensuremath{5.2}}
\def\valIso    {\ensuremath{-0.43^{+0.25}_{-0.22}\pm0.10}\xspace}
\def\valIsoOm    {\ensuremath{-0.49^{+0.30}_{-0.27}\pm0.10}\xspace}
\def\valYrp    {\ensuremath{23.3^{+8.1}_{-7.3}\pm3.1}\xspace}
\def\valYrz    {\ensuremath{34.9^{+8.6}_{-7.9}\pm1.2}\xspace}
\def\valYom    {\ensuremath{12.4^{+6.6}_{-5.7}\pm2.0}\xspace}
\def\valBrpBKst {\ensuremath{0.030^{+0.012}_{-0.011}}\xspace}  
\def\valBroBKst {\ensuremath{0.024\pm0.006}\xspace}  
\def\valBomBKst {\ensuremath{0.012^{+0.007}_{-0.006}}\xspace}  
\def\valBrhoOmBKst {\ensuremath{0.039\pm0.008}\xspace}  
\def\valBrhoBKst {\ensuremath{0.042\pm0.009}\xspace}  
\def\valVtdVtsRp {\ensuremath{0.198^{+0.039} _{-0.035} \pm 0.016}\xspace} 
\def\valVtdVtsRz {\ensuremath{0.254^{+0.033} _{-0.031} \pm 0.021}\xspace} 
\def\valVtdVtsOm {\ensuremath{0.202^{+0.058} _{-0.050} \pm 0.016}\xspace} 
\def\valVtdVtsROm {\ensuremath{0.233^{+0.025} _{-0.024}{^{+0.022}_{-0.021}} }\xspace} 
\def\valVtdVtsRho {\ensuremath{0.235^{+0.026} _{-0.025} \pm 0.020}\xspace}

\begin{document}

\preprint{\babar-PUB-\BABARPubYear/\BABARPubNumber}
\preprint{SLAC-PUB-\SLACPubNumber} 

\begin{flushleft}
\babar-PUB-\BABARPubYear/\BABARPubNumber\\
SLAC-PUB-\SLACPubNumber\\
\end{flushleft}

\title{
{\large \bf
Measurements of Branching Fractions for $\brpg$, $\brzg$, and $\bomg$} 
}

\smallskip

%
\author{B.~Aubert}
\author{M.~Bona}
\author{Y.~Karyotakis}
\author{J.~P.~Lees}
\author{V.~Poireau}
\author{E.~Prencipe}
\author{X.~Prudent}
\author{V.~Tisserand}
\affiliation{Laboratoire de Physique des Particules, IN2P3/CNRS et Universit\'e de Savoie, F-74941 Annecy-Le-Vieux, France }
\author{J.~Garra~Tico}
\author{E.~Grauges}
\affiliation{Universitat de Barcelona, Facultat de Fisica, Departament ECM, E-08028 Barcelona, Spain }
% ...
\collaboration{The \babar\ Collaboration}
\noaffiliation

\begin{abstract} 
  We present branching fraction measurements for the radiative decays
  \brpg, \brzg, and \bomg.
  The analysis is based on a data sample of \numBB\ million \BB\
  events collected with the \babar\ detector at the PEP-II
  asymmetric-energy \BF\ located at the Stanford Linear Accelerator
  Center (SLAC). We find $\BR(\brpg) = (\valBFrp)\valBFbase$,
  $\BR(\brzg) = (\valBFrz)\valBFbase$, and a $90\%$ C.L. upper
  limit $\BR(\bomg)< \valBFomUL\valBFbase$, where the first error is
  statistical and the second is systematic. We also measure the
  isospin-violating quantity $\Gamma(\brpg)/2\Gamma(\brzg) - 1 = \valIso$.
\end{abstract}


\maketitle
\section{INTRODUCTION}
Within the standard model (SM), the radiative decays \brpg, \brzg,
and \bomg~\footnote{Charge conjugate modes are implied throughout}
proceed mainly through a \bdg electroweak penguin amplitude with a
virtual top quark in the loop. Hence, the decay rates depend on the
magnitude of the Cabibbo-Kobayashi-Maskawa (CKM) matrix element
$V_{td}$. The branching fraction results from recent next-to-leading
order calculations are listed in Table~\ref{tab:predictions}. 
While these exclusive decay rates have a large theoretical uncertainty
dominated by the imprecise knowledge of the form factors,
some of this uncertainty cancels in the ratio of
\hbox{$B\to\rho(\omega)\gamma$\ to \bkg} branching fractions. 
This ratio provides a constraint on the ratio of the CKM matrix
elements \VtdVts, which can also be obtained from the ratio of \Bd
and \Bs mixing frequencies~\cite{bsmixing}. Physics beyond the SM
could affect differently $B\to\rho(\omega)\gamma$ and \Bd/\Bs mixing,
and hence create inconsistencies between the results obtained from the
two methods. 

The ratio of \hbox{$B\to\rho(\omega)\gamma$ to \bkg} branching
fractions is related to $\VtdVts$~\cite{alivtdvtstheory} via
\begin{align}\label{eq:vtd}
 \frac{{\cal B}[B\to \rho(\omega)\gamma]}{{\cal B}(B \rightarrow
    K^{*}\gamma)} & =  S \left| \frac{V_{td}}{V_{ts}} \right|^{2}
    \nonumber \\
 & \cdot \left(\frac{1-m_{\rho(\omega)}^{2}/m_{B}^{2}}{1-m_{K^{*}}^{2}/m_{B}^{2}}\right)^{3}
  \zeta^{2}_{\rho(\omega)} \left[1+\Delta R_{\rho(\omega)}\right].
\end{align}
The coefficient $S$ is $1$ for $\rho^+$ and $\frac{1}{2}$ for $\rho^0$
or $\omega$, $m$ is the particle mass, $\zeta_{\rho(\omega)}$ is the
ratio of the form factors for the decays $B\to \rho(\omega)\gamma$ and
\bkg, and $\Delta R_{\rho(\omega)}$ accounts for differences in decay
dynamics, including weak annihilation contributions. The precision of
the \VtdVts determination can be improved by using an average
branching fraction for $B\to\rho(\omega)\gamma$ decays. Within the
SM, the isospin asymmetry between \brpg and \brzg is dominated by weak
annihilation contributions, and is expected to be small; on the other
hand, the asymmetry between \brzg and \bomg can be sizable, due to the
difference in the form factors~\cite{ali2006,Ball2007}.   

We report an updated study of the decays \brpg, \brzg, and \bomg
based on \numBB\ million \BB\ events, corresponding to an integrated
luminosity of \lumBB~fb$^{-1}$, a data sample 25\% larger than that
use in our previous publication~\cite{babar07}. In
addition, we reduce backgrounds considerably by using a multivariate
algorithm based on bootstrap-aggregated (bagged) decision trees
(BDTs)~\cite{breiman} and additional discriminating variables to
separate signal from background.
\begin{table}
\centering
\caption{\label{tab:predictions} Recent predictions of the
  branching fractions.}
\renewcommand{\arraystretch}{1.3}
\begin{tabular}{l@{\hspace{0.3cm}}c@{\hspace{0.3cm}}c@{\hspace{0.3cm}}c@{\hspace{0.3cm}}}
\hline
\hline
Mode & \multicolumn{3}{c}{Branching fraction ($\times 10^{-6}$)} \\
     & Ref.~\cite{ali2006} & Ref.~\cite{Bosch2002}  & Ref.~\cite{Ball2007}     \\
\hline
$\brpg$  & $1.41\pm0.27$  & $1.58^{+0.53}_{-0.46}$ &  $1.16\pm0.26$  \\
$\brzg$  & $0.69\pm0.12$  & $0.76^{+0.26}_{-0.23}$ &  $0.55\pm0.13$  \\
$\bomg$  & $0.55\pm0.09$  &                        &  $0.44\pm0.10$  \\
\hline
\hline 
\end{tabular}
\end{table}

\section{The \babar\ DETECTOR and DATA SET}
The data sample is collected with the \babar\ detector at
the \pep\ asymmetric--energy \epem storage ring at a center of mass
(CM) energy near $\sqrt{s} = 10.58$\gev, corresponding to the
$\FourS$ resonance (on-resonance). Charged particle 
trajectories and energy loss (\dedx) are measured with a five-layer silicon
vertex tracker (SVT) and a 40-layer drift chamber (DCH) in a 1.5 T
magnetic field. Photons and electrons are detected in 
a CsI(Tl) crystal electromagnetic calorimeter (EMC) with photon energy
resolution $\sigma_E / E = 0.023 (E/\mathrm{GeV})^{-1/4} \oplus
0.019$. A ring-imaging Cherenkov detector based on the detection of
internally reflected Cherenkov light (DIRC) provides information for
charged particle identification. The $K$-$\pi$
separation in the DIRC is above 4$\sigma$ at laboratory momenta
up to 3\GeVc. In order to identify muons, the magnetic flux return is
instrumented with resistive plate chambers and limited streamer
tubes. A detailed description of the detector can be found
elsewhere~\cite{ref:babar}.

We use a GEANT4-based~\cite{ref:GEANT4} Monte Carlo (MC) simulation to 
model the \babar\ detector response, taking into account the varying
accelerator and detector conditions. Dedicated signal and background
MC samples are used to optimize selection criteria, to obtain signal
efficiencies and to validate the analysis. Data control samples,
including 41~fb$^{-1}$ of data collected about 40 MeV below the \BB 
production threshold (off-resonance), are used to study backgrounds
coming from continuum $\ep\en \to \qqbar$, with $\q=\u,\d,\s,\c$.

\section{EVENT RECONSTRUCTION AND BACKGROUND SUPPRESSION} 
The decays \broomg are reconstructed by combining a high-energy photon
with a vector meson reconstructed in the decay modes \rzpipi, \rppipi,
and \ompipi. The dominant source of background is coming from
continuum events that contain a high-energy photon from \piz or \etaz
decays or from initial-state radiation (ISR). There are also
significant backgrounds from \B meson decays. The decays \bkg,
$K^{*}\to K\pi$, can mimic the signal when the kaon is misidentified
as a pion. Decays of $\B\to(\rho/\omega)(\piz/\eta)$ with a
high-energy photon from the \piz or $\eta$ decay also mimic the
signal. In addition, there are other \B backgrounds originating mainly
from $B\to X_s\gamma$ and $B\to X(\piz/\eta)$ decays.

The event selection and background suppression are performed in two
steps. We apply a set of loose selection criteria to select
well-measured photons and charged pions and to reject background
events that are kinematically very different from the signal
events. For events that pass the loose event selection criteria, we 
then use the BDT technique to further reduce background.

\subsection{LOOSE SELECTION}
We reduce background contributions from continuum processes by
considering only events for which the ratio $R_2$ of second-to-zeroth
Fox-Wolfram moments~\cite{fox}, calculated using the momenta of all
charged and neutral particles in the event, is less than 0.7.

A photon candidate is identified as a cluster of energy deposited in
contiguous EMC crystals, and not associated with any charged track.   
The high energy photon must have energy $1.5<E_{\g}<4.4\GeV$ in the
laboratory frame and $1.5<E^*_\g <3.5\GeV$ in the CM frame, be
well-contained within the EMC acceptance with polar angle
$-0.74<\cos\theta<0.93$, and be isolated by at least 25 cm at the
entrance of the EMC from any other photon candidate or charged track. 
The distribution of the energy deposition is required to be
consistent with that of a photon shower. 

\begin{figure*}
\includegraphics[width=0.4\linewidth,clip=true]{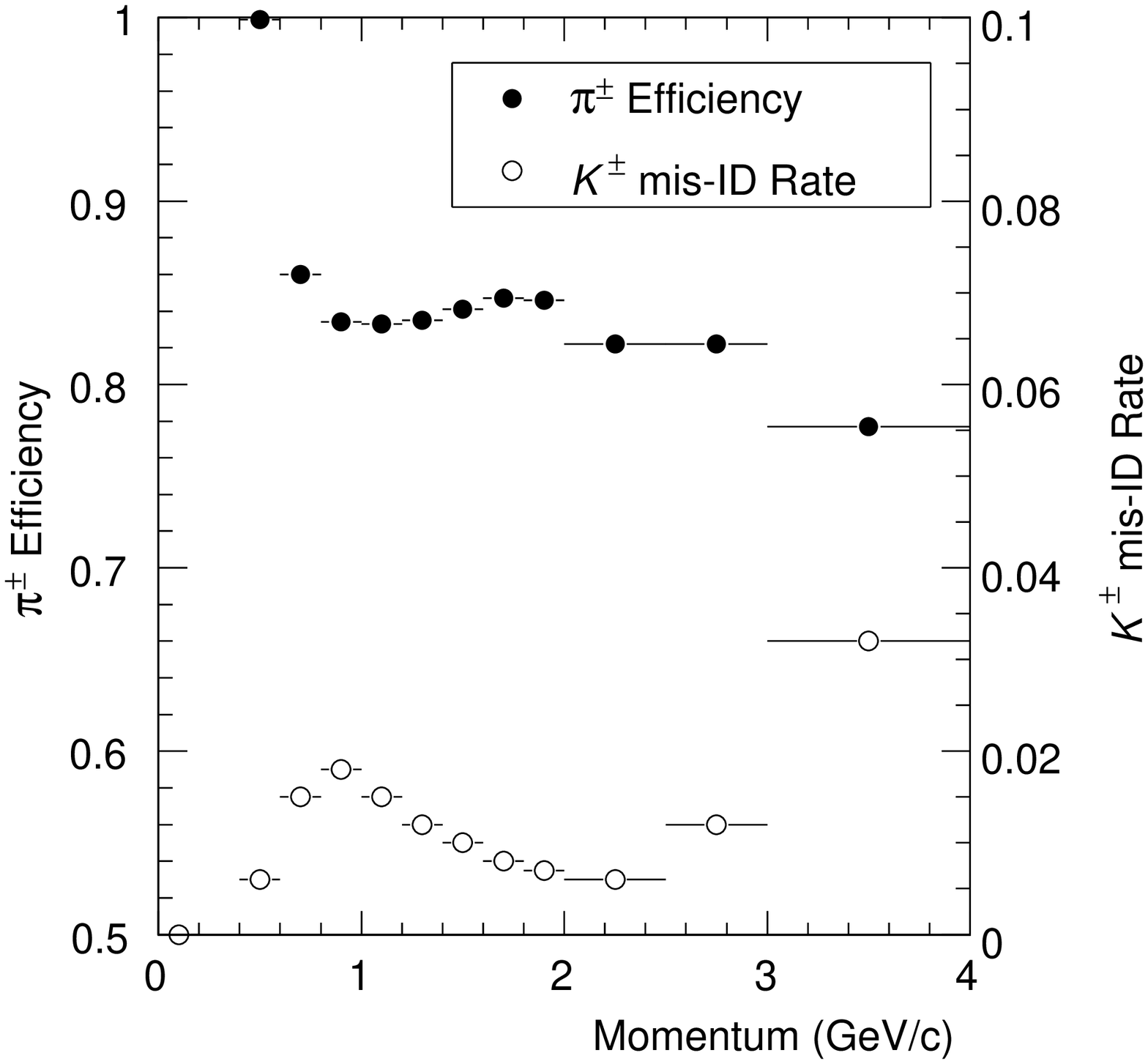}
\hspace{10pt}
\includegraphics[width=0.4\linewidth,clip=true]{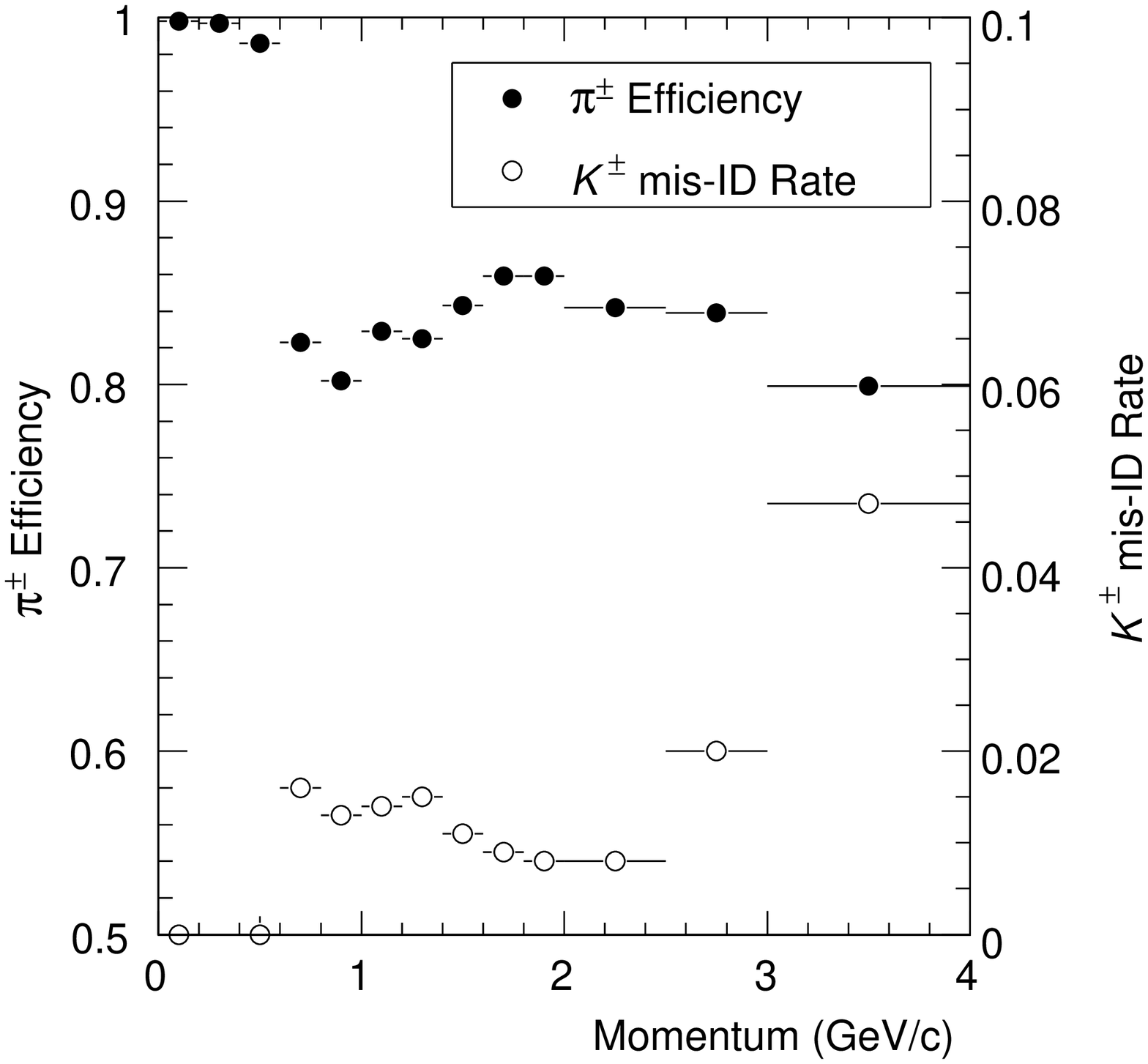}\\
\caption{Performance of the charged-pion identification
  requirement applied to \brg decays, evaluated using the $D^*$
  control sample. Filled circles are for $\pi^{\pm}$ efficiency and
  use the left-hand scale. Open circles are for $K^{\pm}$
  mis-identification and use the right-hand scale. The plot on the
  left shows results for continuum MC events and the plot on the right
  shows results for data. }
\label{fig:pid} 
\end{figure*}

Charged-pion candidates are selected from well-reconstructed tracks
that have at least 12 DCH hits used in the track fit and a minimum
momentum transverse to the beam direction of $100\MeVc$. The tracks
are required to originate near the interaction point (IP): the
distance of closest approach to the IP must be less than 10 cm along the
beam direction and less than 2 cm in the plane perpendicular to the
beam direction. The \pipm identification is based on a likelihood
$L_i$ computed for particle hypothesis $i (=\pi, K, p)$ using 
\dedx measured in the SVT and DCH and the information of Cherenkov
photons detected by the DIRC. The selection criteria are optimized to
reject charged kaons produced in \bkg decays. The pion candidates 
in \bomg must have $L_K/(L_K+L_\pi)<0.5$ and $L_p/(L_p+L_\pi)<0.98$
and must not be consistent with being an electron. The pion 
candidates in \brg must have $L_K/(L_K+L_\pi)<0.2$ and
$L_p/(L_p+L_\pi)<0.5$ and must not be consistent with either an
electron or a muon candidate hypothesis; in addition, for all
candidates with laboratory 
momenta above $0.6\GeVc$, the number of photons observed in the DIRC
is required to be consistent with the number that is expected for the
pion hypothesis. The performance of the pion identification
requirements is evaluated with the decay $D^{*+}\to D^0(\to
K^-\pi^+)\pi^+$, which provides a large, clean sample of 
\pipm and $K^{\pm}$. Using the results shown in Fig.~\ref{fig:pid}, we
find the pion identification requirement retains $85\%$ of the pions
from \brg decays and rejects $99\%$ of the kaons from \bkg decays.

We form \piz candidates from pairs of photons with energies greater
than $50\mev$ in the laboratory frame and an invariant mass
$m_{\g\g}$ in the range $115-150\MeVcc$. 
We combine the identified pions into vector-meson candidates requiring
$630 < m_{\pip\pim} < 960\MeVcc$, $640 < m_{\pip\piz} < 930\MeVcc$, and
$760 < m_{\pip\pim\piz} < 790\MeVcc$ for $\rho^0$, $\rho^+$,
and $\omega$, respectively. The charged-pion pairs are required to
originate from a common vertex. 

The photon and $\rho/\omega$ candidates are combined to form the \B meson
candidates. We define $\de \equiv E^*_{B}-\sqrt{s}/2$, where
$E^*_B = E^*_{\rho/\omega}+E^*_{\g}$ is the CM energy of the $B$ meson
candidate. The \de distributions of signal events are expected to peak near
zero with a resolution of about $50\MeV$ dominated by the photon energy
resolution, and to have a tail in the negative region due to photon
energy loss in the detector. We also define the
beam-energy-substituted mass $\mes \equiv \sqrt{
  s/4-{\mathbf{p'}}_{B}^{\;*2}}$, where ${\mathbf{p'}}_B^{\;*}$
is the CM momentum of the \B candidate modified by scaling the photon 
momentum so that $E^*_{\rho/\omega}+{E'}^*_{\g} - \sqrt{s}/2=0$.
This procedure improves the \mes resolution for the signal events in
the \de negative tail. Signal events are expected to have an \mes
distribution centered at the mass of the \B meson $m\rm_B$ with a
resolution of $3\MeVcc$. We consider candidates with
$\mes>5.22\GeVcc$ and $-0.3<\de<0.3\GeV$ for further analysis. This
region includes sidebands that allow the continuum background yields
to be extracted from a fit to the data.

\subsection{BAGGED DECISION TREE}
The bagged decision trees are
trained separately for the \brpg, \brzg, and \bomg channels with MC
simulated signal and background event samples. The background sample
consists of a \BB MC sample that is about 3 times larger than the data 
and of a continuum MC sample is about 1.5 times larger. 
For the input classifiers, we choose approximately
sixty event quantities that characterize the kinematics of
the \piz candidates, the high-energy photon, the vector meson, the \B
meson and the rest of event (ROE), which are the particles that are
not used to reconstruct the \B candidate. These quantities all have
distributions that agree well between off-resonance data and continuum
MC events.

To reduce combinatorial background in the reconstructed \piz
candidates, we use in the BDT the invariant mass $m_{\g\g}$ and
$\cos\theta_{\g\g}$, the cosine of the opening angle between the
photons in the laboratory frame.

We associate the high-energy photon candidate $\g$ with each of the
other photons $\g'$ in the event and calculate the likelihood ratio
\begin{equation}
\L_i = \frac{{\cal P}(m_{\g\g'},E_{\g'} | i ) }
     {{\cal P}(m_{\g\g'},E_{\g'} | \mbox{signal}
       )+{\cal P}(m_{\g\g'},E_{\g'} | i ) },
\end{equation}
where $i=\piz,\eta$ and ${\cal P}$ is the probability density function
(PDF) defined in terms of the energy of the second photon in the
laboratory frame $E_{\g'}$ and the invariant mass of the pair
$m_{\g\g'}$. The PDFs are determined from simulated signal and
continuum background events. The likelihood ratios $\L_{\piz}$ and
$\L_{\etaz}$ are used in the BDT to reject high energy photons from
\piz and \etaz decays. 

To reject background events from $\B\to\rho(\piz/\eta)$ and
$\B\to\omega(\piz/\eta)$, we also use the vector meson helicity
angle, $\theta_H$, which is defined as the angle between the $B$
momentum vector and the $\pi^+$ track calculated in the $\rho$ rest
frame for a $\rho$ meson, or the angle between the $B$ momentum vector
and the normal to the $\omega$ decay plane for an $\omega$ meson. This
variable is useful because in signal events, the vector meson is
transversely polarized, while in the background events it is 
longitudinally polarized. 

Variables used in the BDT to reduce continuum background include
$R_2$, the significance of the separation of the two \B vertices along
the beam axis ($S_{\Delta{z}}$), the polar angle of the \B candidate
momentum in the CM frame with respect to the beam axis ($\theta_B^*$),
and $R_{2}'$, which is $R_2$ in the frame recoiling against the photon
momentum. We compute the moments $M_i\equiv \sum_j p^*_j\cdot
|{\cos}\theta^*_j|^i/\sum_j p^*_j$ with $i=1,2,3$, where $p^*_j$
is the momentum of each particle $j$ in the ROE and $\theta^*_j$ is
the angle of the momentum with respect to an axis. We use the $M_i$
with respect to the photon direction and the ROE thrust axis. We also
include flavor-tagging variables~\cite{babartag} to exploit the
differences in lepton and kaon production between background and \B
decays.

\begin{figure}
\includegraphics[width=0.8\linewidth,clip=true]{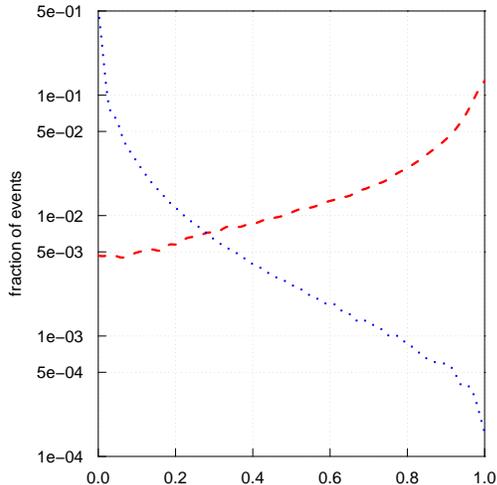}\\
\caption{Distributions of the BDT output for \brzg in signal (dashed)
  and background (dotted) MC samples. The distributions are normalized
  to 1.}
\label{fig:bdt} 
\end{figure}

While we find that all the variables contribute to the sensitivity of
the analysis, the most effective ones are $S_{\Delta{z}}$,
$\cos\theta_{\g\g}$, $R_2$, $\cos\theta_B^*$, $M_3$ with respect to
the photon direction, the missing mass of the ROE, $\cos\theta_H$, and
$\L_{\piz,\eta}$. The distribution of the BDT output for the decay
\brzg is shown in Fig.~\ref{fig:bdt}. We require the BDT output to be
greater than $0.94$ (0.93) for \brg (\bomg). These selection
requirements have been optimized for maximum statistical signal
significance with assumed signal branching fractions of
$1.0$\valBFbase and $0.5$\valBFbase for the charged and neutral modes,
respectively. The signal significance is determined from a fit
described in the next section. For the signal events that
pass the loose selection criteria, the BDT requirements have an
efficiency of 19\% for \brpg, 31\% for \brzg, and 34\% for \bomg.

In events where multiple candidates are present, we select the one
with the reconstructed vector meson mass closest to the nominal
mass. This criteria is chosen because the mass of the vector meson is
found to be uncorrelated with the variables used in the fit. 
After applying all the selection criteria described above to signal MC 
samples, we find signal efficiencies of \effrp\% for \brpg, \effrz\%
for \brzg, and \effom\% for \bomg (taking into account the branching
fraction ${\cal B}(\ompipi)=0.892\pm0.007$~\cite{pdg}),
while backgrounds are reduced by $O(10^{-5})$. 

\section{MAXIMUM LIKELIHOOD FIT}
We determine signal yields from an unbinned maximum likelihood fit to
\mes and \de. The likelihood function for a signal mode $k$ ($=\rhop\g$, 
$\rhoz\g$, $\omega\g$) with a sample of $N_k$ events is
defined as 
\begin{equation}
  {\cal L}_{k}=\exp{\left(-\sum_{i=1}^{N_{\mathrm{hyp}}} n_{i}\right)}
  \left
      [\prod_{j = 1}^{N_k}\left(\sum_{i=1}^{N_{\mathrm{hyp}}} n_i{\cal P}_{i}(\vec{x}_j;\vec{\alpha}_i)\right)\right],
      \label{LH}
\end{equation}
where $N_{\mathrm{hyp}}$ is the number of event hypotheses,
and $n_i$ is the yield for each.  For \bomg, three event hypotheses
are considered: signal, continuum background and combinatorial \B backgrounds. For
\brzg, a \bkzg\ background hypothesis is also included, while for \brpg,
a combined \bkpg/\rppiz hypothesis is included. Since the correlations
between \mes and \de are found to be negligible in MC event samples,
we define the probability density function $\mathcal{P}_{i}(\vec{x_j};
\vec{\alpha_{i}})$ as the product of individual PDFs for each
observable ${x}_{j}$={\mes, \de} given the set of parameters
$\vec{\alpha}_{i}$.

The individual PDFs are determined from fits to dedicated MC event
samples. The signal \mes PDFs are parametrized by a Crystal Ball (CB)
function~\cite{CryBall} and the \de PDFs are parametrized as 
\begin{equation}
f(x) =
\exp \left(\frac{-(x-\mu)^2} {2 \sigma^2_{L,R}+\alpha_{L,R} (x-\mu)^2}
\right),
\end{equation}
 where $\mu$ is the peak position of the distribution,
$\sigma_{L}$ and $\sigma_{R}$ are the widths on the left and right of the peak, and
$\alpha_{L}$ and $\alpha_{R}$ are a measure of the tails on the left and right of the
peak, respectively. The peak positions and widths of the signal \mes
and \de PDFs are corrected for the observed difference between data
and MC samples of \bkg decays. 
The PDFs for the remaining \bkzg\ and combined \bkpg/\brpiz
backgrounds are determined from dedicated MC samples that are 100
times larger than the data. These PDFs are described by a CB function
for \mes, with a peak position the same as that of the signal PDF but
having a much larger width, and a CB function for \de, with a peak
position near $-80\MeV$. The negative \de peak position is due either
to a kaon misidentified as a pion in \bkg\ or to a single missing
photon in \brpiz. The \mes and \de PDFs for all other \B backgrounds
are determined from the \BB MC sample. The \mes spectra peak slightly
in the signal region, and therefore are parametrized by a CB function,
while the \de spectra are parametrized by an exponential function. The
continuum \mes and \de PDFs are parametrized by an ARGUS threshold
function~\cite{Argus} and a first order polynomial, respectively. 

The fit to the data determines the signal yield $n_{\rm{sig}}$, the
continuum yield and the shape parameters of the continuum PDFs. The
shape parameters of the signal and \B background PDFs are fixed in the
fit. The relative yield between the peaking and the other \B 
backgrounds is fixed to the value obtained from known branching
fractions~\cite{pdg} and selection efficiencies determined from MC
event samples. The overall yields of the \B backgrounds are also fixed.
All fixed parameters are later varied to evaluate systematic errors in
$n_{\rm{sig}}$. 

We validate the fitting procedure using ensembles of signal and
background events. Two types of ensembles are produced: both
signal and background events generated using the PDFs described
above; signal events randomly sampled from the GEANT4 MC events and
background events generated using the corresponding PDFs. No bias
is found in the fit to these event samples. 

\begin{figure*}
\includegraphics[width=0.45\linewidth,clip=true]{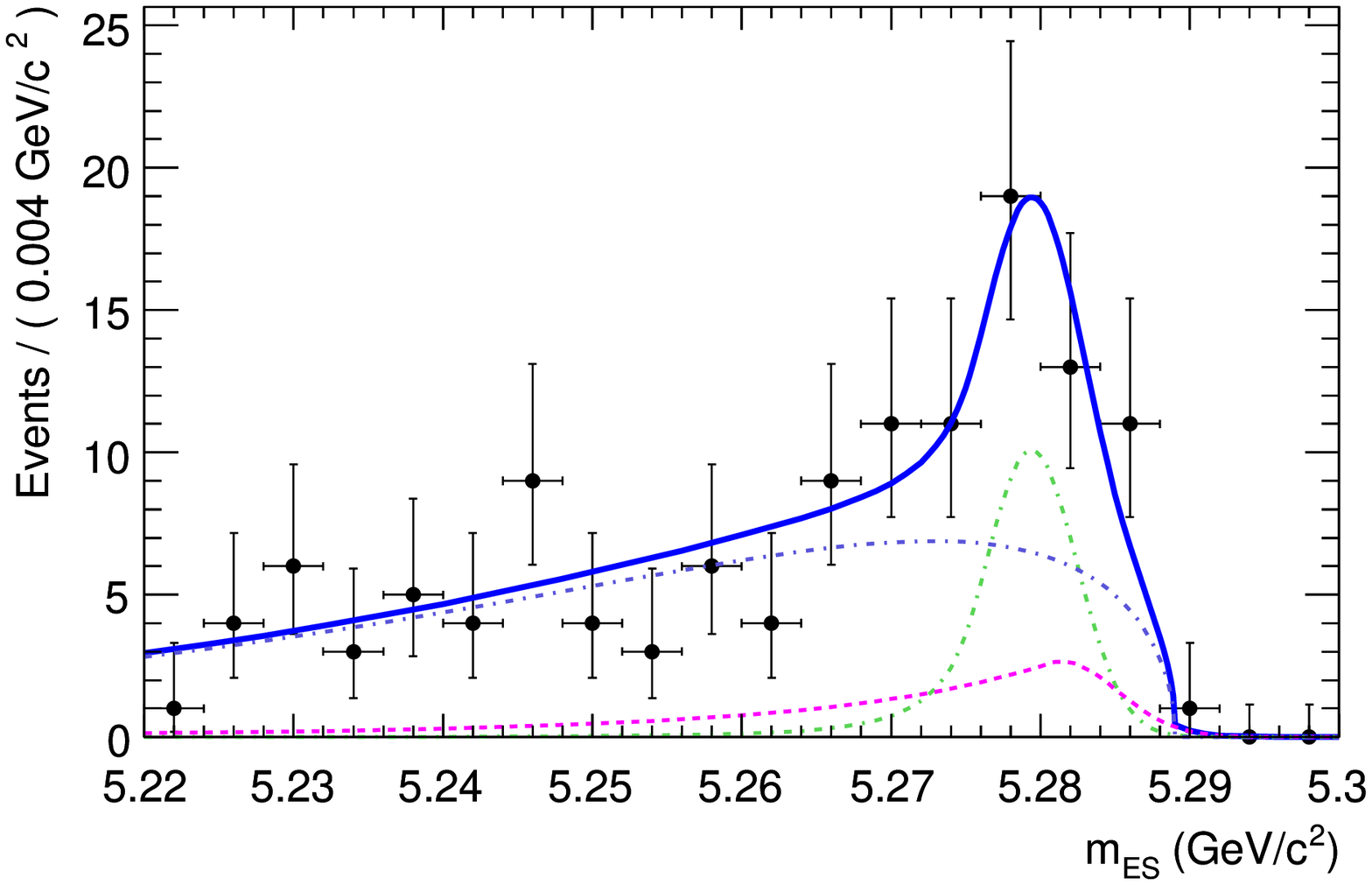}%
\includegraphics[width=0.45\linewidth,clip=true]{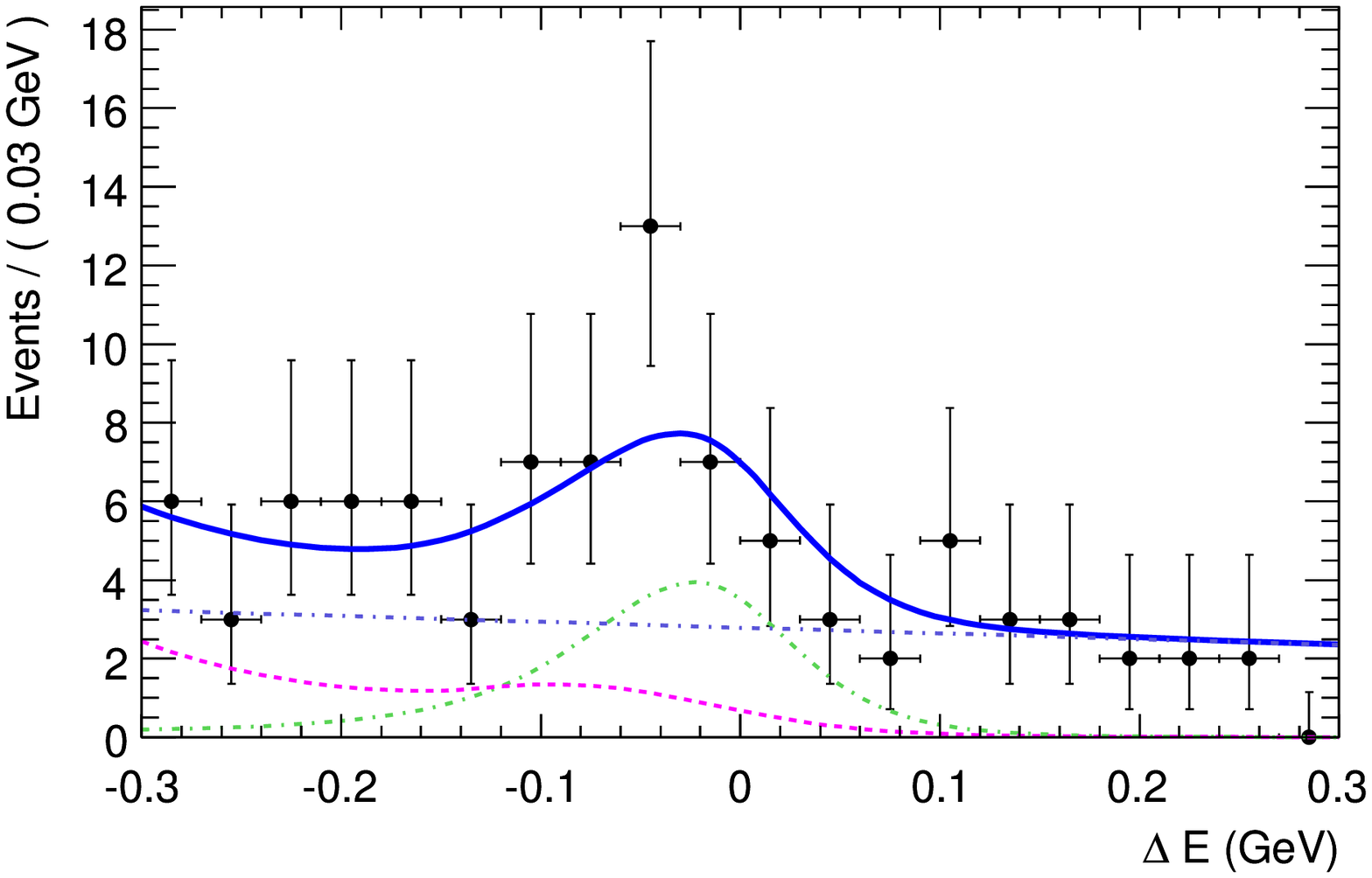}\\
\includegraphics[width=0.45\linewidth,clip=true]{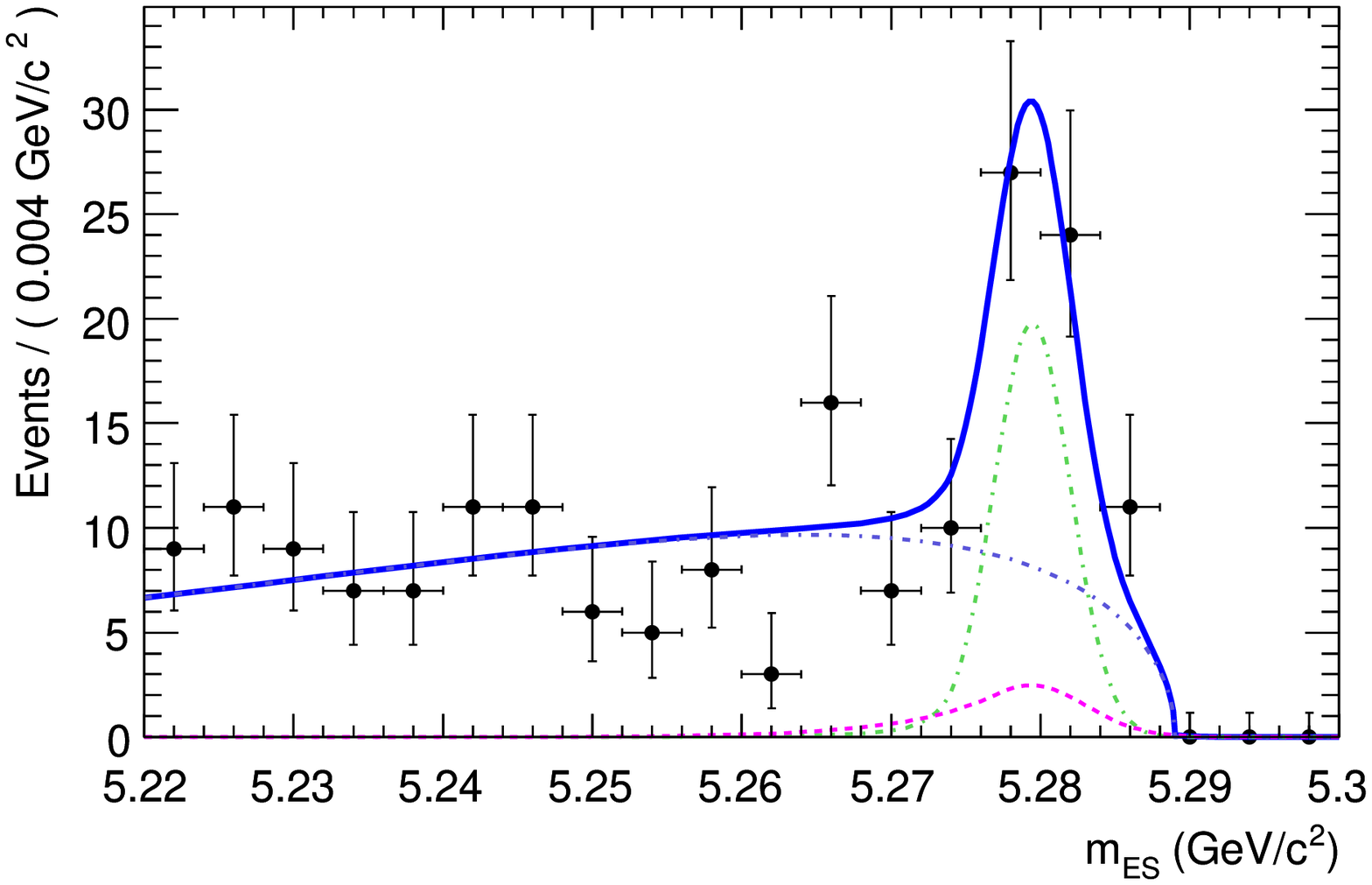}%
\includegraphics[width=0.45\linewidth,clip=true]{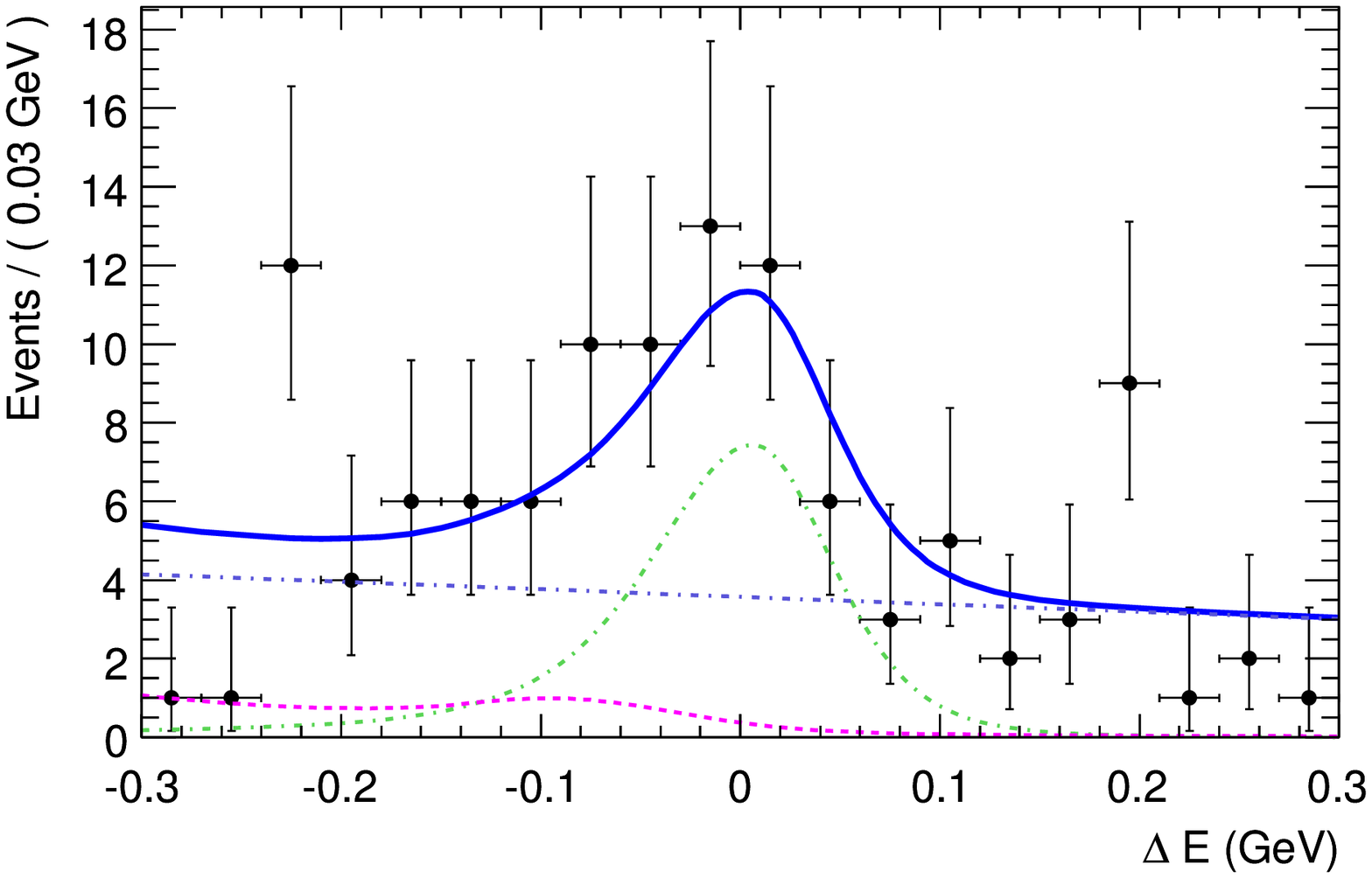}\\
\includegraphics[width=0.45\linewidth,clip=true]{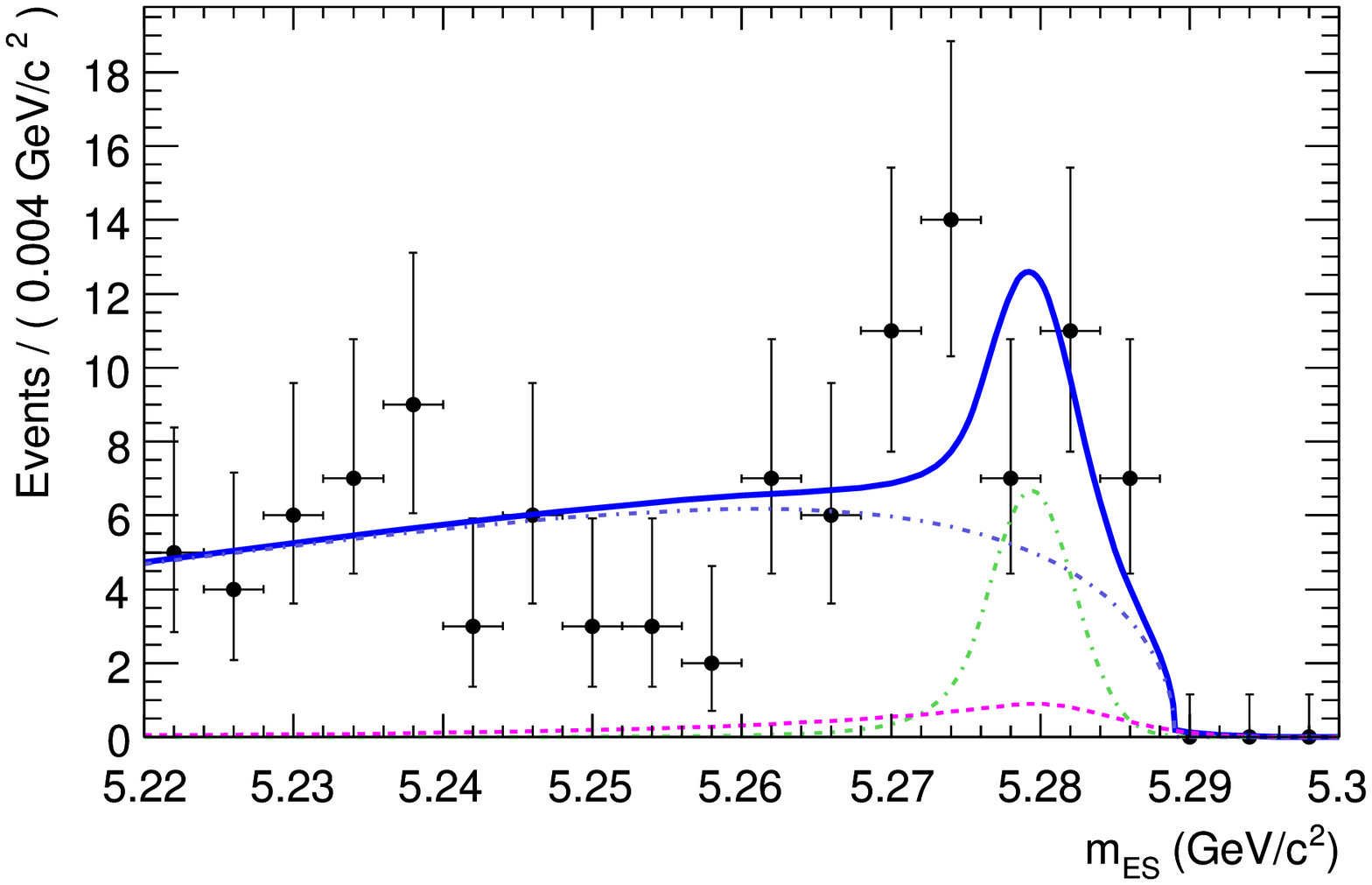}%
\includegraphics[width=0.45\linewidth,clip=true]{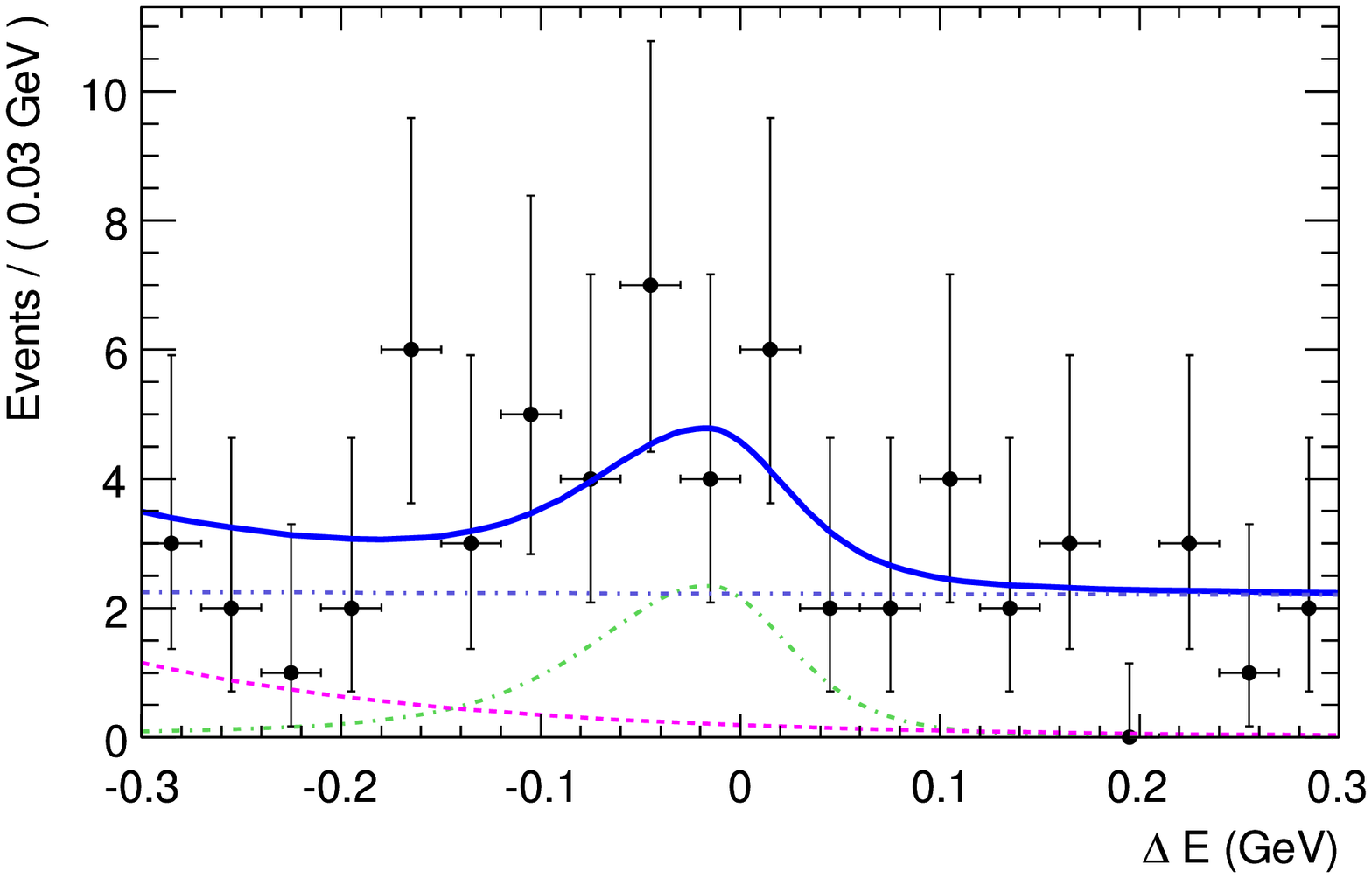}\\
\caption{\de and \mes projections of the fits for the decay modes
  $\brpg$ (top), $\brzg$ (middle), and $\bomg$ (bottom). For
  illustrative purpose only, these plots are made by requiring
  $-0.2<\de<0.1\GeV$ for the \mes projections and $\mes>5.27\GeVcc$
  for the \de projections. The points are data, the solid line is the
  total PDF, the dashed line is the sum of \B background PDFs, the
  dash-dotted line is the continuum PDF, and the dotted line is the
  signal PDF. }\label{fig:rhoChfit}
\end{figure*}

Figure \ref{fig:rhoChfit} shows the data points and the projections of
the fit results for \de and  \mes separately for each decay mode.
The signal yields are reported in Table~\ref{tab:results}. The
significance is computed as $\sqrt{2\Delta\ln\mathcal{L}}$, where
$\Delta\ln\mathcal{L}$ is the log-likelihood difference between the
best fit and the null-signal hypothesis. To take into account the
systematic error in $n_{\rm{sig}}$, the likelihood function is
convolved with a Gaussian distribution that has a width equal to the
systematic error. 
%

\section{SYSTEMATIC UNCERTAINTIES}
Table~\ref{tab:syst} gives the contributions
to the systematic uncertainties. The systematic error affecting on the
signal efficiency includes uncertainties on tracking, particle
identification, \g and \piz reconstruction, and BDT selection. 
The modeling of signal and background in the fit
contributes to the uncertainties on the signal yields. 
\begin{table}
\centering
\caption{\label{tab:results} The signal yield $n_{\mathrm{sig}}$,
significance $\Sigma$ in standard deviations including the systematic
error in $n_{\mathrm{sig}}$, efficiency $\epsilon$, and branching
fraction $\mathcal{B}$ for each mode. The first error is statistical
and the second is systematic. The branching fractions for
\bromg and \brg are obtained with the assumption of isospin and
$\rm{SU}(3)_F$ symmetries.}
\renewcommand{\arraystretch}{1.3}
\begin{tabular}{l@{\hspace{0.3cm}}c@{\hspace{0.3cm}}c@{\hspace{0.3cm}}c@{\hspace{0.3cm}}c@{\hspace{0.3cm}}c}
\hline
\hline
Mode & $n_{\mathrm{sig}}$ & $\Sigma$
     & $\epsilon (\%)$    & $\mathcal{\B} (10^{-6})$ \\
\hline
$\brpg$  & $\valYrp$      & $3.2\sigma$ &  $\effrp$ & $\valBFrp$ \\
$\brzg$  & $\valYrz$      & $5.4\sigma$ &  $\effrz$ & $\valBFrz$ \\
$\bomg$  & $\valYom$      & $2.2\sigma$ &  $\effom$ & $\valBFom$ \\
\hline
$\bromg $ & & $6.5\sigma$ &  & $\valBFav $ \\
\hline
$\brg $ & & $6.0\sigma$ &  & $\valBFavrho $ \\
\hline
\hline
\end{tabular}
\end{table}
\begin{table}
\renewcommand{\arraystretch}{1.3}
\centering \caption{\label{tab:syst}
Fractional systematic errors (in \%) of the measured branching fractions.}
\begin{tabular*}{\linewidth}{lp{0.9cm}p{0.9cm}p{0.9cm}p{0.65cm}p{0.9cm}}
\hline \hline
\multicolumn{1}{p{3.5cm}}{Source of error}
&  $\rho^{+}\g$  &  $\rho^{0}\g$
&  $\omega\g$  & $\rho\g$ &  $(\rho/\omega)\g$    \\
\hline
Tracking efficiency             &     0.4  &  0.4  &   0.4  &  0.4  &  \multicolumn{1}{c}{0.4}  \\ 
Particle identification         &     1.0  &  2.0  &   1.0  &  1.4  &  \multicolumn{1}{c}{1.2} \\
Photon selection                &     2.8  &  2.8  &   2.8  &  2.8  &  \multicolumn{1}{c}{2.8}  \\ 
$\pi^0$ reconstruction          &     3.0  &  -    &   3.0  &  1.7  &  \multicolumn{1}{c}{2.0} \\
BDT efficiency                  &     9.3  &  4.2  &   5.1  &  7.0  &  \multicolumn{1}{c}{7.5}  \\
Signal model                     &    7.1  &  2.1  &  16.3  &  3.0  &  \multicolumn{1}{c}{3.0} \\
Background model                 &   10.9  &  2.8  &   2.7  &   4.3  &  \multicolumn{1}{c}{3.6} \\
$B\overline{B}$ counting         &     1.1  &  1.1  &  1.1  &  1.1  &  \multicolumn{1}{c}{1.1} \\
\BR($\omega\to\pi^+\pi^-\pi^0$) &     -    &  -    &   0.8  &  -    &  \multicolumn{1}{c}{0.1} \\
\hline
Sum in quadrature               &     16.7  & 6.6  &   17.9  &   9.5  &  \multicolumn{1}{c}{9.5} \\
\hline \hline
\end{tabular*}
\end{table}

The errors in BDT selections are determined from a control sample of
the decay $B^0 \to K^{*0}(\to K^+\pi^-)\gamma$ for the $\rho^0$ mode
and a sample of $B^+ \to K^{*+}(\to K^+\pi^0)\gamma$ for the $\rho^+$
and $\omega$. These \bkg decays are
kinematically similar to \brog decays. The events
are required to pass all applicable loose selection criteria, except
the pion identification requirements. We also require the invariant
mass $0.80<m_{K^+\pi^-}<1.0\GeVcc$ and $0.82<m_{K^+\pi^0}<0.96\GeVcc$. 
The BDT output classifiers are computed from the decision trees
trained for the corresponding signal modes. The differences in the BDT 
selection efficiencies between the \bkg data and MC samples are used
to correct the signal efficiencies. The efficiency correction factor
is $0.88\pm0.09$ for \brpg, $0.91\pm0.04$ for \brzg and
$0.90\pm0.05$ for \bomg. 
The uncertainty of the correction is taken as the systematic
error. The large BDT systematic error for the decay \brpg is due to
the limited size of the $B^+ \to K^{*+}(\to K^+\pi^0)\gamma$ sample. As
a means of validating the BDT technique, we apply the same analysis
technique to the \bkg data control samples and measure 
the branching fractions for \bkg. The results are consistent with the
world averages~\cite{pdg}.

The error in the pion identification requirements is estimated
using the $D^*$ control sample as shown in Fig.~\ref{fig:pid}.
Based on the difference of a momentum-weighted efficiency between the
continuum MC sample and data, a $1\%$ systematic error per
charged-pion is assigned to the \brg decays. The MC sample is in
better agreement with data for the looser pion identification criteria applied to \bomg and a $0.5\%$ error per charge-pion is assigned.
The uncertainties from tracking, \piz reconstruction, and photon
selection are also determined from suitable independent data control
samples.

To estimate the uncertainty related to the modeling of the signal and
background, we vary the parameters of the PDFs that are fixed in the fit
within their errors. We vary the relative and absolute
normalizations of \B background components that are fixed in
the fit based on a kaon mis-identification study using the $D^*$
control sample as shown in Fig.~\ref{fig:pid}. We find the difference
in the momentum-weighted kaon mis-identification rates between the
data and MC samples is $23\%$ and conservatively vary the \bkg
background yield by $30\%$. The effect of the uncertainty of
$\BR(\brppiz)$~\cite{pdg} is also considered for the decay \brpg. For
all the variations, the corresponding changes in the extracted signal
yield are taken as systematic uncertainties, which are then combined,
taking into account correlations. The error on background modeling for
\brpg is dominated by uncertainties in \B background PDFs. 

\section{RESULTS}
To calculate the branching fractions from the measured signal yields,
we assume $\BR(\Upsilon (4S)\to\BzBzb) = \BR(\Upsilon (4S)\to\BpBm) = 0.5$.
The results are listed in Table~\ref{tab:results}.
For $\bomg$, we also compute the $90\%$ confidence level
(C.L.) upper limit $\BR(\bomg) < \valBFomUL\valBFbase$
using a Bayesian technique,
assuming a prior flat in the branching fraction
and taking into account the systematic uncertainty.

We test the hypothesis of isospin symmetry by measuring
the quantity
\begin{eqnarray*}\label{eq:isos}
\Delta_{\rho}=\frac{\Gamma(\brpg)}{2\Gamma(\brzg)} - 1 =
\valIso.
\end{eqnarray*}
Most theoretical calculations~\cite{ali2006,Bosch2002,Ball2007,Lu2005}
predict small $\Delta_{\rho}$. For example the estimate
in Ref.~\cite{Ball2007} is $-0.05\pm0.03$ for $\gamma=60\degrees$ and
$-0.10\pm0.02$ for $\gamma=70\degrees$, where $\gamma$ is
the phase of $V^{*}_{ub}$. Our result is consistent with these
predictions within the large experimental errors. However, it is worth
noting that a recent calculation~\cite{Kim2008} indicates that
nonperturbative charming penguin contributions can accommodate large
$\Delta_{\rho}$. We also measure the $\rm{SU}(3)_F$-violating quantity
\begin{eqnarray*}\label{eq:su3}
\Delta_{\omega}=\frac{\Gamma(\bomg)}{\Gamma(\brzg)} - 1 =
\valIsoOm,
\end{eqnarray*}
which is consistent with the theoretical calculations.

We extract average branching fractions using a simultaneous fit to
all the relevant decay modes with the constraints on the widths of the
decay modes: $\Gamma_{\brpg}=2\Gamma_{\brzg}=2\Gamma_{\bomg}$.
The average branching fractions are defined as
\begin{equation}
  \BR(\brg) \equiv \frac{1}{2} \Bigl[ \BR (\brpg) 
      +  2\frac{\tau_{\Bp}}{\tau_{\Bz}} \BR(\brzg) \Bigr] \\
\end{equation}
and 
\begin{align}
   \avbr \equiv \frac{1}{2} \Bigl\{ \BR(\brpg) 
  \;\;\;\;\;\;\;\;\;\;\; 
  \;\;\;\;\;\;\;\;\;\;\; \nonumber\\
    +  \frac{\tau_{\Bp}}{\tau_{\Bz}}\left[ 
      \BR(\brzg) + \BR(\bomg) \right] \Bigr\},
\end{align}
where $\tau_{\Bp}/\tau_{\Bz}$ is the measured ratio between the
charged and neutral \B meson lifetimes, for which the current world
average is $1.071\pm0.009$~\cite{pdg}. 
Our measurements of the individual branching fractions are consistent
with this hypothesis, with a $\chi^2$ of 2.3 for 2 degrees of freedom. 
We find:
\begin{eqnarray*}\label{eq:res1}
  \BR(\brg) &=& (\valBFavrho )\valBFbase \\
  \avbr &=& (\valBFav) \valBFbase.
\end{eqnarray*}

Using the world average value of $\BR(\bkpg)=(4.03\pm0.26)\times
10^{-5}$, $\BR(\bkzg)=(4.01\pm0.2)\times 10^{-5}$~\cite{pdg}, and the
isospin averaged branching fraction $\BR(\bkg)=(4.16\pm0.17)\times
10^{-5}$, we calculate
\begin{eqnarray*}\label{eq:rhokst}
  R_{\rhop} &=& \frac{\BR(\brpg)}{\BR(\bkpg)} = \valBrpBKst \\
  R_{\rhoz} &=&   \frac{\BR(\brzg)}{\BR(\bkzg)} = \valBroBKst \\
  R_{\omega} &=&   \frac{\BR(\bomg)}{\BR(\bkzg)} = \valBomBKst \\
  R_{\rho} &=&   \frac{\BR(\brg)}{\BR(\bkg)} = \valBrhoBKst \\
  R_{\rho/\omega} &=&   \frac{\avbr}{\BR(\bkg)} = \valBrhoOmBKst.
\end{eqnarray*}
These ratios of branching fractions can be used to calculate
\VtdVts~\cite{alivtdvtstheory, Bosch:2004nd, Ball2007}. 
Following Eq. 1 and using $1/\zeta_{\rho}=1.17\pm0.09$,
$1/\zeta_{\omega}=1.30\pm0.10$~\cite{Ball2007}, $\Delta
R_{\rhop}=0.057^{+0.057}_{-0.055}$, $\Delta
R_{\rhoz}=0.006^{+0.046}_{-0.043}$, and
$\Delta R_{\omega}=-0.002^{+0.046}_{-0.043}$~\cite{ali2006} ,
we obtain
\begin{eqnarray*}\label{eq:res5} 
 \VtdVts_{\rhop} &=& \valVtdVtsRp \\
 \VtdVts_{\rhoz} &=& \valVtdVtsRz \\
 \VtdVts_{\omega} &=& \valVtdVtsOm,
\end{eqnarray*}
where the first error is experimental and the second is theoretical. 
Using the average branching fractions and following Ref.~\cite{ali2006},
we obtain 
\begin{eqnarray*}\label{eq:res6} 
  \VtdVts_{\rho} &=& \valVtdVtsRho \\
  \VtdVts_{\rho/\omega} &=& \valVtdVtsROm.
\end{eqnarray*}
Similar values are found following Ref.~\cite{Ball2007}. These results are
consistent with the value of this 
ratio, $0.208\pm0.002$(exp)$^{+0.008}_{-0.006}$(theory)~\cite{pdg},
obtained from the studies of \Bd and \Bs mixing by the CDF and D0
Collaborations.

\section{SUMMARY}
We report the updated measurements of the branching fractions for the
radiative decays \brpg, \brzg, and \bomg
\begin{eqnarray*}
\BR(\brpg) &=& (\valBFrp)\valBFbase\\
\BR(\brzg) &=& (\valBFrz)\valBFbase\\
\BR(\bomg) &<& \valBFomUL\valBFbase\;(90\%\;{\rm{C.L.}}).
\end{eqnarray*}
We test the hypothesis of isospin symmetry by measuring the quantity
$\Delta_{\rho} = \valIso$. We also measure the averaged
branching fractions $ \BR(\brg) = (\valBFavrho )\valBFbase $ and
$\avbr= (\valBFav)\valBFbase$.  These results are in 
good agreement with, and supersede, the previous published \babar\
measurement~\cite{babar07}, which uses a subsample of the data used
for this analysis. These results are also consistent with the
measurements from Belle~\cite{belle}.
These branching fraction measurements are used to extract \VtdVts
in a way that is complementary to the approach using \B
mixing~\cite{bsmixing}.\\

\section{ACKNOWLEDGMENTS}
We are grateful for the 
extraordinary contributions of our \pep2\ colleagues in
achieving the excellent luminosity and machine conditions
that have made this work possible.
The success of this project also relies critically on the 
expertise and dedication of the computing organizations that 
support \babar.
The collaborating institutions wish to thank 
SLAC for its support and the kind hospitality extended to them. 
This work is supported by the
US Department of Energy
and National Science Foundation, the
Natural Sciences and Engineering Research Council (Canada),
the Commissariat \`a l'Energie Atomique and
Institut National de Physique Nucl\'eaire et de Physique des Particules
(France), the
Bundesministerium f\"ur Bildung und Forschung and
Deutsche Forschungsgemeinschaft
(Germany), the
Istituto Nazionale di Fisica Nucleare (Italy),
the Foundation for Fundamental Research on Matter (The Netherlands),
the Research Council of Norway, the
Ministry of Education and Science of the Russian Federation, 
Ministerio de Educaci\'on y Ciencia (Spain), and the
Science and Technology Facilities Council (United Kingdom).
Individuals have received support from 
the Marie-Curie IEF program (European Union) and
the A. P. Sloan Foundation.


\begin{thebibliography}{99}
\bibitem{ali2006}
  A.~Ali and A.~Y.~Parkhomenko, arXiv:hep-ph/0610149; 
  updated analysis (to be published).

\bibitem{Bosch2002}
  S.~W.~Bosch and G.~Buchalla, \npb{621}, 459 (2002).

\bibitem{Ball2007}
  P.~Ball and R.~Zwicky, \jhep{0604}, 046 (2006);
  P.~Ball, G.~Jones and R.~Zwicky, \jprd{75} 054004 (2007).

\bibitem{bsmixing}
  A.~Abulencia {\it et al.}, [CDF Collaboration],
  \jprl{97}, 242003 (2006);
  V.~Abazov {\it et al.}, [D0 Collaboration], 
  \jprl{97}, 021802 (2006).

\bibitem{alivtdvtstheory}
  A.~Ali and A.~Y.~Parkhomenko, \epjc{23}, 89 (2002);
  A.~Ali, E.~Lunghi, and A.~Y.~Parkhomenko, \plb{595}, 323 (2004).
      
\bibitem{babar07}
  B.~Aubert {\it et al.},   [\babar\ Collaboration],
  \jprl{98}, 151802 (2007).

\bibitem{breiman}
  L.~Breiman, 
  \ml{24(2)}, 123 (1996).
 
\bibitem{ref:babar}
  B.~Aubert {\it et al.},  [\babar\ Collaboration],
  \nima{479}, 1 (2002).
  
\bibitem{ref:GEANT4}
  S.~Agostinelli {\it et al.} [GEANT4 Collaboration]
  \nima{506}, 250 (2003).

\bibitem{fox} 
  G.~C.~Fox and S.~Wolfram, 
  \npb{149}, 413 (1979).
  
\bibitem{babartag}
  B.~Aubert {\it et al.}, [\babar\ Collaboration],
  \jprl{89}, 201802 (2002).
  
\bibitem{pdg}
  W.~M.~Yao {\it et al.}  [Particle Data Group], 
  \jpg{33}, 1 (2006) and 2007 partial update for 2008.

\bibitem{CryBall}
  J.~E.~Gaiser {\it et al.} [Crystal Ball Collaboration], \jprd{34}, 711 (1986).
    
\bibitem{Argus}
  H.~Albrecht {\it et al.}, [ARGUS Collaboration], \zpc{48}, 543 (1990).

\bibitem{Lu2005}
  C.-D.~Lu {\it et al.},
  \jprd{72} 094005 (2005).

\bibitem{Kim2008}
  C.~Kim, A.~K.~Leibovich and T.~Mehen, arXiv:0805.1735 (hep-ph).

\bibitem{Bosch:2004nd}
  S.~W.~Bosch and G.~Buchalla,
  \jhep{0501}, 035 (2005). 

\bibitem{belle}
  D.~Mohapatra {\it et al.},  [Belle Collaboration]
  \jprl{96}, 221601 (2006);
  N.~Taniguchi {\it et al.}, [Belle Collaboration]
  arxiv:0804.4770 (hep-ex).

\end{thebibliography}
\end{document}